\documentstyle[12pt]{article}
\begin{document}

\author{M. E. Palistrant}
\title{Comment on "Theory of two-band superconductors"}
\maketitle

{\it Institute of Applied Physics, Chishinau, Moldova.} \newline
\newline

The main stages of development of the theory of superconducting systems with
overlapping energy bands are formulated.

The main references on the classical papers of the author of this theory,
Prof. V.A. Moskalenko, and his coworkers are listed. The list also includes
papers related to high - temperature superconductivity. Some peculiarities
of the two - band model, which give qualitatively new results in comparison
to the usual one - band model, are enumerated.

We also list our own publications on thermodynamics and collective
oscillations in the two - band systems which are based on a generalization
of Moskalenko model of systems with overlapping energy bands for the case of
a reduced density of charge carriers.

The application of two-band model for the description of the thermodynamical
properties of the compound $MgB_2$ is also discussed.\newline
\vspace{5mm}

The model of a superconductor with the overlapping of energy bands on the
Fermi surface has been  proposed by Prof. V. Moskalenko more than 40 years
ago \cite{Moskalenko} and some time later by H. Suhl, B. Matthias and L.
Walker [2]. On the basis of this model Prof. Moskalenko with coworkers
(Institute of Applied Physics Moldavian Academy of Sciences) have
investigated the thermodynamic and electromagnetic properties of multi-band
superconductors. A few books and a lot of articles on this problem had been
published, for instance, \cite{Moskalenko_1}-\cite{Moskalenko_10}.

After the discovery of high temperature superconductivity we attempted to
apply our theory to the explanation of a number of properties of these
materials. For example, a review of our results has been published in 1991
\cite{Moskalenko_11}.

Later we have adapted the generalized Moskalenko model for the investigation
of thermodynamic properties and collective oscillations in multi-band
systems with a reduced density of charge carries (see, for instance, \cite%
{Palistrant}-\cite{Palistrant_4}).

Recently, many experimental results have been reported on the
superconductivity in $MgB_{2},$ confirming the theoretical concept of the
two - band superconductivity.

Consideration of the overlapping of energy bands leads not only to
quantitatively but, in some cases, to qualitatively new results as compared
to the case of one-band superconductors. For example:

1) In the two-band systems a high temperature superconductivity is possible
not only in the case of an attractive interaction between electrons, but
even if the interaction has a repulsive character ($\lambda _{nm}<0,n;m=1-2$%
), provided the relation $\lambda _{11}\lambda _{22}-\lambda _{12}\lambda
_{21}<0$ is fulfilled \cite{Palistrant}

2. In the impurity doped two-band superconductors, e.g., the Anderson
theorem is violated, namely, a dependence of thermodynamic quantities on
concentrations of a non-magnetic impurity appears due to the interband
scattering of electrons on impurity atoms \cite{Moskalenko_2}.

3.  In the two-band superconductors the temperature dependence of the upper
critical field $H_{c2}$ in the vicinity of $T_{c}$ has a positive curvature
due to the relation $v_{F_{1}}\not=v_{F_{2}}$ ($v_{Fn}$ - the velocity of
electron on $n$-th cavity of Fermi-surface). In the one-band case the
curvature is negative. \cite{Moskalenko_11}.

4. In a two-band superconductor an exciton-type collective oscillation mode
(the Leggett mode) appears as a consequence of fluctuations of the phase of
order parameters in different bands. In a three-band system (and also in a
two-band one with a reduced density of charge carriers allowing to map it
onto a three-band model) there can be two such oscillatory modes \cite%
{Kochorbe_1}, \cite{Kochorbe_2}.

5. On the basis of the theory of superconductivity with overlapping energy
bands one can explain a large number of experimental results on High - $T_{c}
$ materials \cite{Moskalenko_11}.

6.The interest to the two - band model for superconductivity has recently
been renewed after the discovery the superconductivity in $MgB_{2}$ $%
(T_{c}\sim 40K)$. A lot of properties of the two-band superconductors were
rediscovered in the very recent investigations on diborides and
borocarbides. Our review article, published in {\it Usp. Fiz. Nauk.} [13].
gives a complete list of the classical achievements in the problem.

The theory presented in Refs. \cite{Moskalenko}-\cite{Moskalenko_11}
contains the main physical concepts, the basic equations of the model and
analytic expressions for the thermodynamic and electromagnetic
characteristics of a superconductor with overlapping energy bands on the
Fermi surface. The obtained results are valid for arbitrary values of
parameters of the adiabatic two-band model (for pure and doped systems).
Therefore, this theory is applicable to the description of the thermodynamic
and electromagnetic properties of the two-band superconductors characterized
by some peculiarities which impose restrictions on the parameters of the
theory. In our opinion, this corresponds, in particular, to the
superconducting compound $MgB_{2}$.

As follows from the overview of experimental and theoretical investigations
of thermodynamic properties of $MgB_{2}$ \cite{Canfield}, \cite{Bouquet},
this compound is a many-band superconductor and anomalies of its properties
are captured by the two-band theory of pure \cite{Shulga} and doped
superconductors \cite{Golubov}, \cite{Brinkman}. As a matter of fact, this
theory includes the same peculiarities as proposed by Moskalenko and his
coworkers many years ago (see \cite{Moskalenko_11} and references therein).
In particular, for the ratios of the order parameters in different bands ($%
\Delta _{1}$ and $\Delta _{2}$) and the superconducting transition
temperature ( $T_{c}$ ) we have $\frac{2\Delta _{1}}{T_{c}}>3.5$;\thinspace\
$\frac{2\Delta _{2}}{T_{c}}<3.5$, while for the relative jump of the heat
capacity $\frac{C_{s}-C_{n}}{C_{n}}<1.43$ irrespectively of the choice of
the model parameters.

We note in addition some other results which were obtained long time ago
\cite{Moskalenko_11} for the two-band model: the possibility of positive
curvature of a function $H_{c2}(T)$ near values $T\sim T_{c}$ due to the
relation {\bf $v_{F1}\not=v_{F2}$ } ($v_{Fn}$ is the electron velocity on
the cavity $n$ of the Fermi surface).

Another important result, which is used in \cite{Bouquet}, \cite{Golubov},
\cite{Brinkman} for the elucidation of properties of $MgB_{2}$ - is the
violation of Anderson theorem and appearance of the $T_{c}$ dependence on
concentration of the non-magnetic impurity due to interband electron
scattering on impurity, absence of the critical concentration of impurities.
This result was obtained by our group in 1965 \cite{Moskalenko_2}.

In our opinion, the essentially new result in the recent development of the
theory of systems with overlapping energy bands is the consideration of the
angular dependence of the energy gap. Such anisotropic gap is observed $%
MgB_{2}$.

In this respect we mention the papers \cite{Mishonov}-\cite{Mishonov_4},
which have developed the method of calculating the heat capacity in the
two-band model taking into account the energy gap anisotropy. The results
obtained the above authors are in good agreement with experimental data on
heat capacity in $MgB_{2}$ of ordered and non-ordered samples.

In our opinion, the description of physical properties of multi-band
superconductors and its application to particular materials ( such as, e.g.,
$MgB_{2}$ ) should include the earlier achievements mentioned above. We
would like to mention the important contribution of Prof. T. Mishonov and
his colleagues, who have taken an objective attitude to the papers published
in 1959 - 1991 years by Prof. V. Moskalenko and his cowokers.


\begin{thebibliography}{99}
\bibitem{Moskalenko} V. A. Moskalenko, {\it Fiz. Met. Metallov} {\bf 8}, 503
(1959); {\it Phys. Met. and Metallov.} {\bf 8}, 25 (1959). % (1)

\bibitem{Suhl} H. Suhl, B. T. Matthias, L. R. Walker, {\it Phys. Rev. Lett.}
{\bf 3}, 552 (1959). % (2)

\bibitem{Moskalenko_1} V. A. Moskalenko and M. E. Palistrant, {\it Docl.
Akad. Nauk SSSR }{\bf 162}, 539 (1965); {\it Sov. Phys. Dokl.} {\bf 10}, 457
(1965). % (3)

\bibitem{Moskalenko_2} V. A. Moskalenko and M. E. Palistrant, {\it Zh. Exsp.
Teor. Fiz. }{\bf 49}, 770 (1965); {\it Sov. Phys. JETP} {\bf 22}, 536
(1965). % (4)

\bibitem{Moskalenko_3} V. A. Moskalenko, L. Z. Kon and M. E. Palistrant,{\it %
Proceedings of Low Temperature Conference LT - X}, Moscow, 181 (1967).
% (5)

\bibitem{Moskalenko_4} V. A. Moskalenko, L. Z. Kon {\it Zh. Exp. Teor. Fiz.}
{\bf 50}, 724 (1966). % (6)

\bibitem{Moskalenko_5} V. A. Moskalenko, {\it Fiz. Met. Metaloved.} {\bf 23}%
, 585 (1967). % (7)

\bibitem{Moskalenko_6} V. A. Moskalenko and M. E. Palistrant, {\it %
Statistical Physics and Quantum Field Theory} [in Russian], Nauka, M. p.262
(1973). % (8)

\bibitem{Moskalenko_7} V. A. Moskalenko, L. Z. Kon and M. E. Palistrant,
{\it Low - Temperature Properties of Metals with Band Spectrum Singularities
}[in Russian], Stiinta, Kishinev (1989). % (9)

\bibitem{Moskalenko_8} V. A. Moskalenko, {\it Zh.Exp. Teor. Fiz}. {\bf 51},
1163 (1966); {\it Sov. Phys. JETP }{\bf 24} 780 (1966). % (10)

\bibitem{Moskalenko_9} V. A. Moskalenko, {\it Electromagnetic and Kinetic
Properties of Superconducting Alloys with Overlapping Energy Bands}[in
Russian], Stiintsa, Kishinev (1976). % (11)

\bibitem{Moskalenko_10} V. A. Moskalenko, {\it Methods of Studying
Electronic Densities of States in Superconducting Alloys}[in Russian],
Stiintsa, Kishinev (1974). % (12)

\bibitem{Moskalenko_11} V. A. Moskalenko, M. E. Palistrant and V. M.
Vacalyuk,{\it Usp. Fiz. Nauk}, {\bf 161}, 155 (1991); {\it Sov. Phys. Usp.}
{\bf 34}, 717 (1991). % (13)

\bibitem{Palistrant} M. E. Palistrant and F. G. Kochorbe, {\it Physica C}
{\bf 194}, 351 - 356 (1992). % (14)

\bibitem{Kochorbe} F. G. Kochorbe and M. E. Palistrant, {\it Zh. Exsp. Teor.
Fiz.} {\bf 104}, 3084 (1993); {\it JETP} {\bf 77}, 442 (1993); {\it Teor.
Mat. Fiz.} {\bf 96}, 459 (1993); {\it Theoret. Mathem. Phys.} {\bf 96}, 1083
(1993) % (15)

\bibitem{Palistrant_1} M. E. Palistrant, {\it Teoret. Mat. Fiz.} {\bf 105},
491 (1995); {\it Theoret. Mathem. Phys.} {\bf 105}, 1593, (1995). % (16)

\bibitem{Palistrant_2} M. E. Palistrant, {\it Teoret. Mat. Fiz.} {\bf 109},
137 (1996); {\it Theoret. Mathem. Phys.} {\bf 109}, 1352, (1996). % (17)

\bibitem{Palistrant_3} M. E. Palistrant, {\it Superconductivity} {\bf 10},
19 (1997). % (18)

\bibitem{Kochorbe_1} F. G. Kochorbe and M. E. Palistrant, {\it Physica C}
{\bf 298}, 217 (1997). % (19)

\bibitem{Kochorbe_2} F. G. Kochorbe and M. E. Palistrant, {\it Zh. Exsp.
Teor. Fiz.} {\bf 114}, 1047 (1998); {\it JETP} {\bf 87}, 570 (1998). % (20)

\bibitem{Palistrant_4} M. E. Palistrant and F. G. Kochorbe, {\it J. Low
Temp. Phys. }{\bf 26}, 299 (2000). % (21)

\bibitem{Canfield} P. C. Canfield, S. L. Bud'ko, D. K. Finnemore, {\it %
Physica C} {\bf 385}, 1 (2003). % (22)

\bibitem{Bouquet} F. Bouquet, J. Wang, I. Sheikin et al., Physica C {\bf 385}%
, 192 (2003). % (23)

\bibitem{Shulga} S. V. Shulga, S. L. Drechsler, G. Fuchs et al.,{\it Phys.
Rev. Lett.} {\bf 80}, 1730 ({\bf 1998}). % (24)

\bibitem{Golubov} A. A. Golubov, L. L. Mazin,{\it Phys. Rev. B} {\bf 55},
15146 (1997)

\bibitem{Brinkman} A. Brinkman, A. A. Golubov, H. Rogalla et al., {\it Phys.
Rev. B} {\bf 65} , 180517 (R) (2002). % (26)

\bibitem{Mishonov} T. Mishonov and E. Penev, arXiv:{\it cond-mat/0206118 V.2
24 June 2002} {\it In. J. Mod. Phys. B} {\bf 16}, 693 (2002).
% (27)

\bibitem{Mishonov_2} T. Mishonov, S. L. Drechsler and E. Penev arxiv: {\it %
cond.-mat./0209192 V.1 8 Sep. 2002}; {\it Mod. Phys. Lett.}, {\bf 17} (2003).
% (28)

\bibitem{Mishonov_3} T. Mishonov, E. Penev, J. O. Indekeu  {\it Phys. Rev B}
{\bf 66}, 066501 (2002)
%(29)

\bibitem{Mishonov_4} T. Mishonov, E. Penev, J. O. Indekeu and V. I.
Pokrovsky, arXiv: {\it cond-mat/0209342 V. 1 14 Sep. 2002.} % (30)
\end{thebibliography}
\end{document}